\documentclass{PoS}

\usepackage{amsmath}

\def\LL{\left}
\def\RR{\right}
\def\beq{\begin{equation}}
\def\eeq{\end{equation}}
\def\bea{\begin{eqnarray}}
\def\eea{\end{eqnarray}}
\def\g5{\gamma_5}

\def\tr{\mathrm{tr}}

\def\Id{\mbox{1\hspace{-1.0mm}I}}

\def\Fig#1{Fig.\,\ref{#1}}

\PoS{PoS(LAT2005)065}

\title{Mass Spectra of Pentaquarks -- Overlap versus Wilson Fermions\thanks{
This work was supported in part by the National Science Council (ROC),
under Grant No. NSC93-2112-M002-016,
and the National Center for High Performance Computation at Hsinchu.}
}
\ShortTitle{Mass Spectra of Pentaquarks}

\author{Ting-Wai Chiu and \speaker{Tung-Han Hsieh}\\
Physics Department, National Taiwan University, Taipei, Taiwan 106, Taiwan \\
	E-mail: \email{twchiu@phys.ntu.edu.tw},
		\email{thhsieh@twcp1.phys.ntu.edu.tw}}

\abstract{%
We investigate the mass spectra of $\Theta^+(udud\bar s)$
in quenched lattice QCD, with overlap and Wilson
fermions respectively. Using three different interpolating operators, 
we measure their correlation matrix, and extract the mass spectra in 
even and odd parity channels, for 100 gauge configurations generated 
with single plaquette action at $\beta=6.1$ on the $20^3 \times 40$ lattice.
The lowest $ 1/2^- $ state agrees with the $ KN $ 
s-wave scattering state, for both fermion schemes. 
On the other hand, for the lowest $ 1/2^+ $ state, 
it is different from any hadron scattering states for the overlap fermion,   
while it seems to agree with $ KN^* $ s-wave for the Wilson fermion.

}

\FullConference{XXIIIrd International Symposium on Lattice Field Theory\\
25-30 July 2005\\
Trinity College, Dublin, Ireland}

\begin{document}

\section{Introduction}

The experimental observation of exotic baryon $\Theta^+(1540)$ (with quantum
numbers of $K^+ n$) by LEPS collaboration at Spring-8 and subsequent
confirmation from some experimental groups has become one of the most
interesting topics in hadron physics. The remarkable features of
$\Theta^+(1540)$ are its strangeness $S=+1$, and its exceptionally narrow
decay width ($<15$ MeV) even though it is about 100 MeV above the $KN$
threshold. Its strangeness $S=+1$ immediately implies that it cannot be an
ordinary 3-quark baryon. Its minimal quark content is $udud\bar s$.
Nevertheless, there are quite a number of experiments which so far have 
not observed $\Theta^+(1540)$. Further, some experiments which had 
claimed positive evidence of $ \Theta $ have switched to negative 
when they attain higher statistics. 
This almost denies the existence of $\Theta^+(1540)$.

To study this problem in lattice QCD, one has to construct
an interpolating operator which has a significant overlap with
the pentaquark state. Then one computes its time-correlation
function to extract the masses of its even and old parity states.
However, any $udud\bar s$ operator must couple to any hadronic states 
with the same quantum numbers (e.g., $KN$ scattering states). To
disentangle the lowest-lying pentaquark state from the various $KN$ 
scattering states, we use three different interpolating operators for
$\Theta^+ (udud\bar s)$ to form a $3\times 3$ correlation matrix,
\beq
C^{\pm}_{ij}(t) = \left< \sum_{\vec x}\tr\Bigl[ \frac{1\pm\gamma_4}{2}
        \left< O_i(\vec x,t)\bar O_j(\vec 0,0)\right>_f \Bigr]\right>_U
\eeq
and extract the masses for both $\pm$ parity states from its
eigenvalues. These three operators are
\bea
\label{eq:O1}
{(O_1)}_{x\alpha} &=&
  [{\bf u}^T C \gamma_5 {\bf d}]_{xc} \
\{ \bar{\bf s}_{x \beta e} (\gamma_5)_{\beta\eta} {\bf u}_{x \eta e}
 (\gamma_5 {\bf d})_{x \alpha c}
- \bar{\bf s}_{x \beta e} (\gamma_5)_{\beta\eta} {\bf d}_{x \eta e} 
(\gamma_5 {\bf u})_{x \alpha c} \}
\\
\label{eq:O2}
{(O_2)}_{x\alpha} &=&
  [{\bf u}^T C \gamma_5 {\bf d}]_{xc} \
\{ \bar{\bf s}_{x \beta e} (\gamma_5)_{\beta\eta} {\bf u}_{x \eta c}
   (\gamma_5 {\bf d})_{x \alpha e}
- \bar{\bf s}_{x \beta e} (\gamma_5)_{\beta\eta} {\bf d}_{x \eta c}
   (\gamma_5 {\bf u})_{x \alpha e} \}
\\
\label{eq:O3}
{(O_3)}_{x\alpha} &=& \epsilon_{cde}
  [{\bf u}^T C \gamma_5 {\bf d}]_{xc} \
  [{\bf u}^T C {\bf d}]_{xd} \ (C \bar {\bf s}^T)_{x\alpha e}
\eea
where $C$ is the charge conjugation which satisfies
$ C \gamma_\mu C^{-1} = -\gamma_\mu^T $ and 
$ (C \gamma_5)^T=-C\gamma_5 $,
and the diquark operator is defined as
$
[{\bf u}^T \Gamma {\bf d} ]_{xa} \equiv \epsilon_{abc} ( 
 {\bf u}_{x\alpha b} \Gamma_{\alpha\beta} {\bf d}_{x\beta c}
-{\bf d}_{x\alpha b} \Gamma_{\alpha\beta} {\bf u}_{x\beta c} )
$, where $\Gamma_{\alpha\beta} = -\Gamma_{\beta\alpha} $.
Thus the diquark transforms like a spin singlet ($1_s$), color anti-triplet
($\bar 3_c $), and flavor anti-triplet ($ \bar 3_f $).
For $ \Gamma = C \gamma_5 $, it transforms as a scalar,
while for $ \Gamma = C $, it transforms like a pseudoscalar.

We generate 100 gauge configurations with Wilson gauge action at $\beta = 6.1$
on the $20^3\times 40$ lattice.  Then we compute the point-to-point quark
propagators for both overlap fermion and Wilson fermion, with both periodic
and anti-periodic boundary conditions in the time direction, respectively.
Then we measure the $3\times 3$ correlation matrix from the {\em averaged}
quark propagators. For
the Wilson fermion, we compute quark propagators for $\kappa =$ 0.149, 0.150,
0.151, 0.152, 0.153, and 0.1534, with stopping criteria $10^{-11}$
for the CG (conjugate gradient) loops. For the overlap fermion, we use the
nested and 2-pass CG algorithms to compute quark propagators. With $m_0=1.3$,
$N_s=128$, and 16 low-lying eigenmodes of $|H_w|$ projected, the Zolotarev
coefficents and poles are fixed with $0.18 \le \lambda(|H_w|) \le 6.3$
for all gauge configurations. We compute quark propagators for 30
bare quark masses in the range $0.03 \le m_q a \le 0.8$, with stopping
criteria $10^{-11}$ and $2\times 10^{-12}$ for the outer and inner CG loops,
respectively. The precision of chiral symmetry and the norm of 
the residue vector for each column of quark
propagators are:
\beq
\sigma = \LL| \frac{Y^{\dagger} S^2 Y}{Y^{\dagger} Y} - 1\RR| < 10^{-14}, 
\hspace{10mm}
|| (D_c + m_q)Y - \Id || < 2\times 10^{-11}
\eeq

\begin{figure}[th]
\begin{center}
\begin{tabular}{@{}c@{}c@{ }c@{}c@{}}
\parbox[b][7.5cm][t]{6.5mm}{(a)} &
\parbox[b][7.5cm][t]{6.5cm}{
        \includegraphics*[height=7cm,width=6cm]{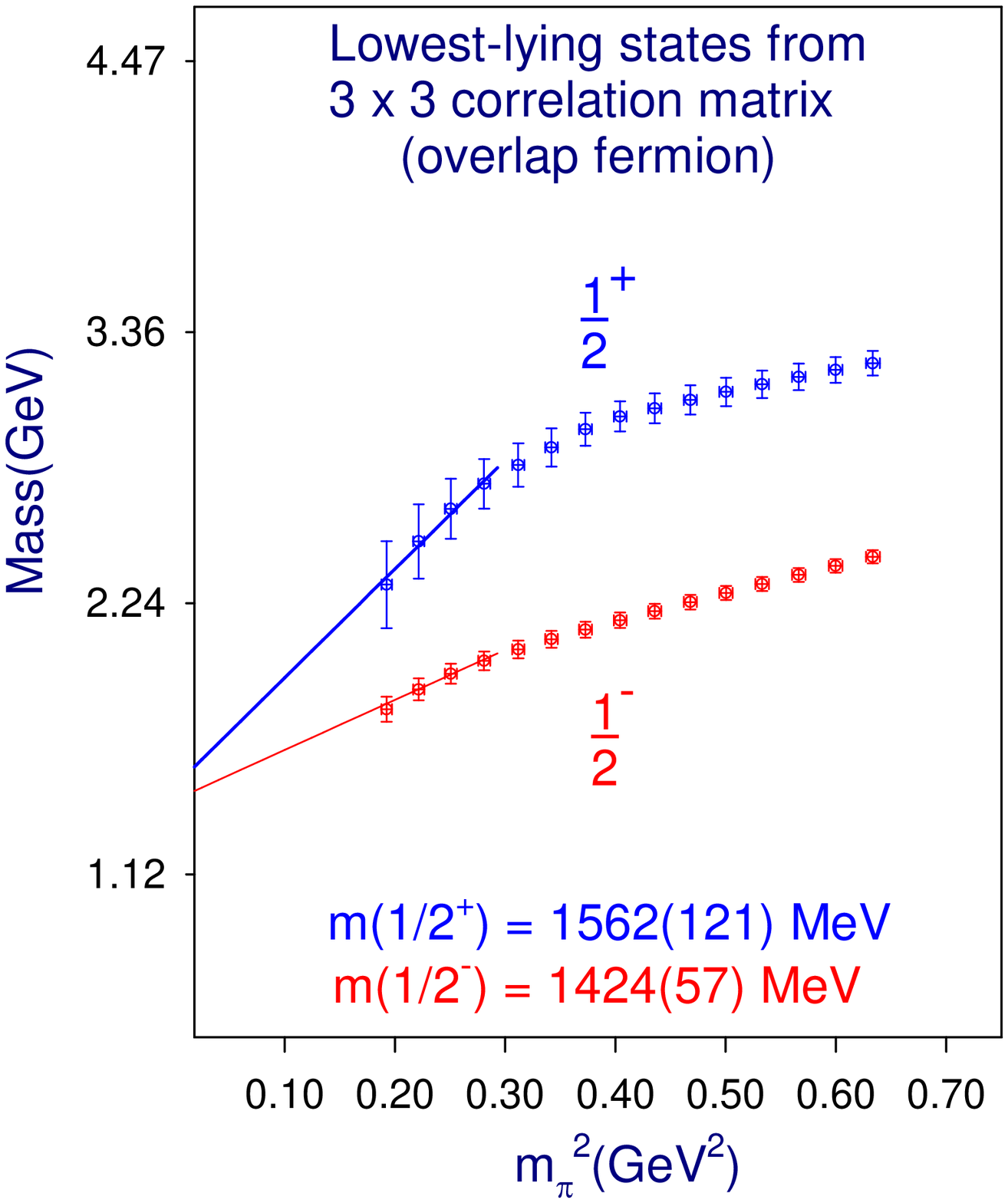}} &
\parbox[b][7.5cm][t]{6.5mm}{(b)} &
\parbox[b][7.5cm][t]{6.5cm}{
        \includegraphics*[height=7cm,width=6cm]{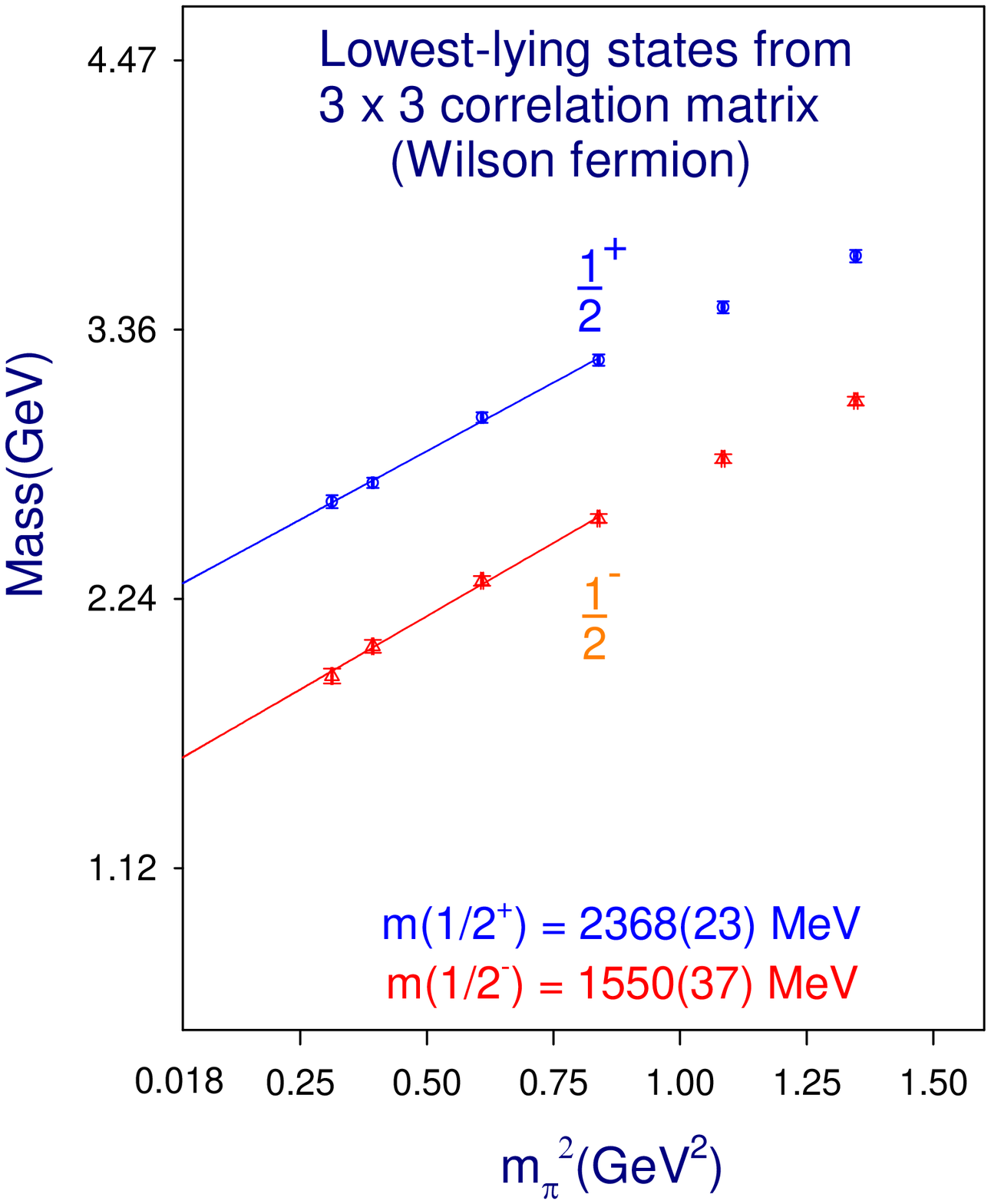}}
\end{tabular}
\vskip -0.8truecm
\caption{Comparision of the lowest-lying states from the $3\times 3$
  correlation matrix, for (a) overlap, and (b) Wilson fermions respectively.}
\label{fig:overlap-wilson-3x3}
\end{center}
\end{figure}

To determine $a^{-1}$, we measure the pion time correlation
function with overlap quark propagators, and extract
pion mass ($m_{\pi}a$) and decay constant
($f_{\pi}a$). With the experimental input $f_{\pi}$ = 131 MeV, we
determine $a^{-1}$ = 2.237(76) GeV.
To determine $m_s$ (for overlap fermion) and $\kappa_s$ (for Wilson fermion),
we extract the vector meson mass from the correlation funtion
\beq
C_V(t) = \frac{1}{3}\sum^3_{\mu=1} \sum_{\vec x}\tr\{
        \gamma_{\mu}(D_c+m_q)^{-1}_{x,0}\gamma_{\mu}(D_c+m_q)^{-1}_{0,x}\}
\eeq
For overlap quark, at $m_q a = 0.08$, it gives
$M_V a = 0.4601(41)$, which amounts $M_V = 1029(10)$ MeV, 
in good agreement with $\phi(1020)$. 
Thus we fix $ m_s a = 0.08 $, and we have
10 quark masses smaller than $m_s$. 
Similarly, for Wilson quark, 
we fix $ \kappa_s = 0.151 $, and 
we have three $ \kappa $'s $ > \kappa_s$.


\section{Results}

In \Fig{fig:overlap-wilson-3x3}, we plot the masses of the lowest-lying
states extracted from the $3\times 3$ correlation matrix, 
for both parity channels, and for overlap and Wilson fermions respectively. 
The solid lines are chiral extrapolation (linear in $m_{\pi}^2$) 
using the smallest four masses. At $ m_\pi = 135 $ MeV, they give 
(a) Overlap fermion: $m(1/2^-) = 1424(57)$ MeV, $ m(1/2^+) = 1562(121)$ MeV; \\
(b) Wilson fermion: $m(1/2^-) = 1550(37)$ MeV, $m(1/2^+) = 2368(23)$ MeV. \\
Evidently, for any parity channel, its behavior  
with the overlap fermion is quite different from that 
with the Wilson fermion. In particular, for the $ 1/2^+ $ state with 
the overlap fermion, 
it experiences a rapid decrease in the regime 
$ m_u a \le 0.045 \simeq 0.56 m_s a $. 
We interpret this as a manifestation of the
{\em diquark correlations} when $m_{u}$ becomes sufficiently small,  
since the overlap fermion preserves the exact chiral symmetry on the 
lattice. On the other hand, Wilson fermion breaks chiral symmetry 
explicitly and the bare quark mass is not well-defined, thus could not 
capture the diquark correlations except in the continuum limit.    
\begin{figure}[th]
\begin{center}
\begin{tabular}{@{}c@{}c@{ }c@{}c@{}}
\parbox[b][7.5cm][t]{6.5mm}{(a)} &
\parbox[b][7.5cm][t]{6.5cm}{
        \includegraphics*[height=7cm,width=6cm]{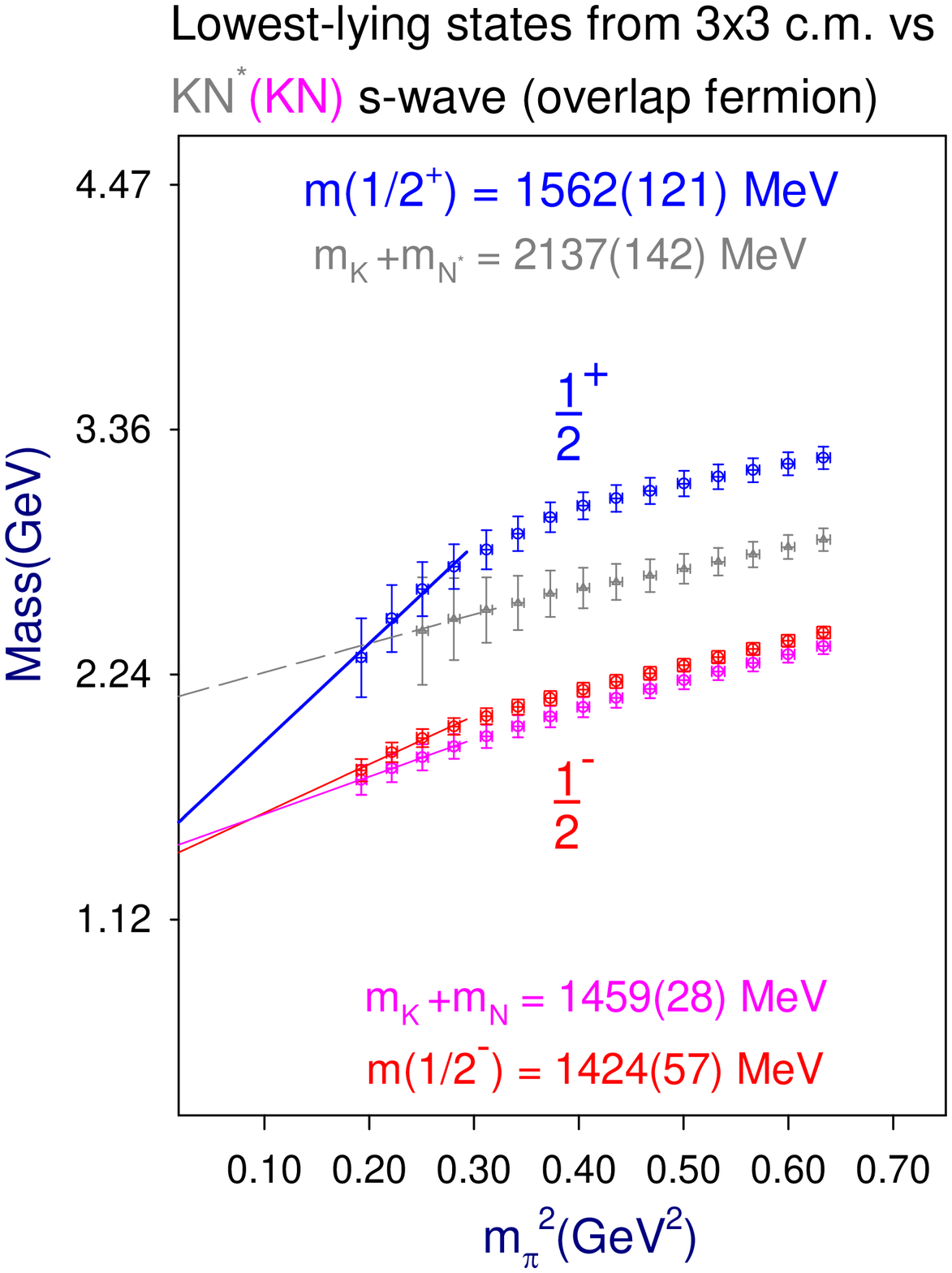}} &
\parbox[b][7.5cm][t]{6.5mm}{(b)} &
\parbox[b][7.5cm][t]{6.5cm}{
        \includegraphics*[height=7cm,width=6cm]{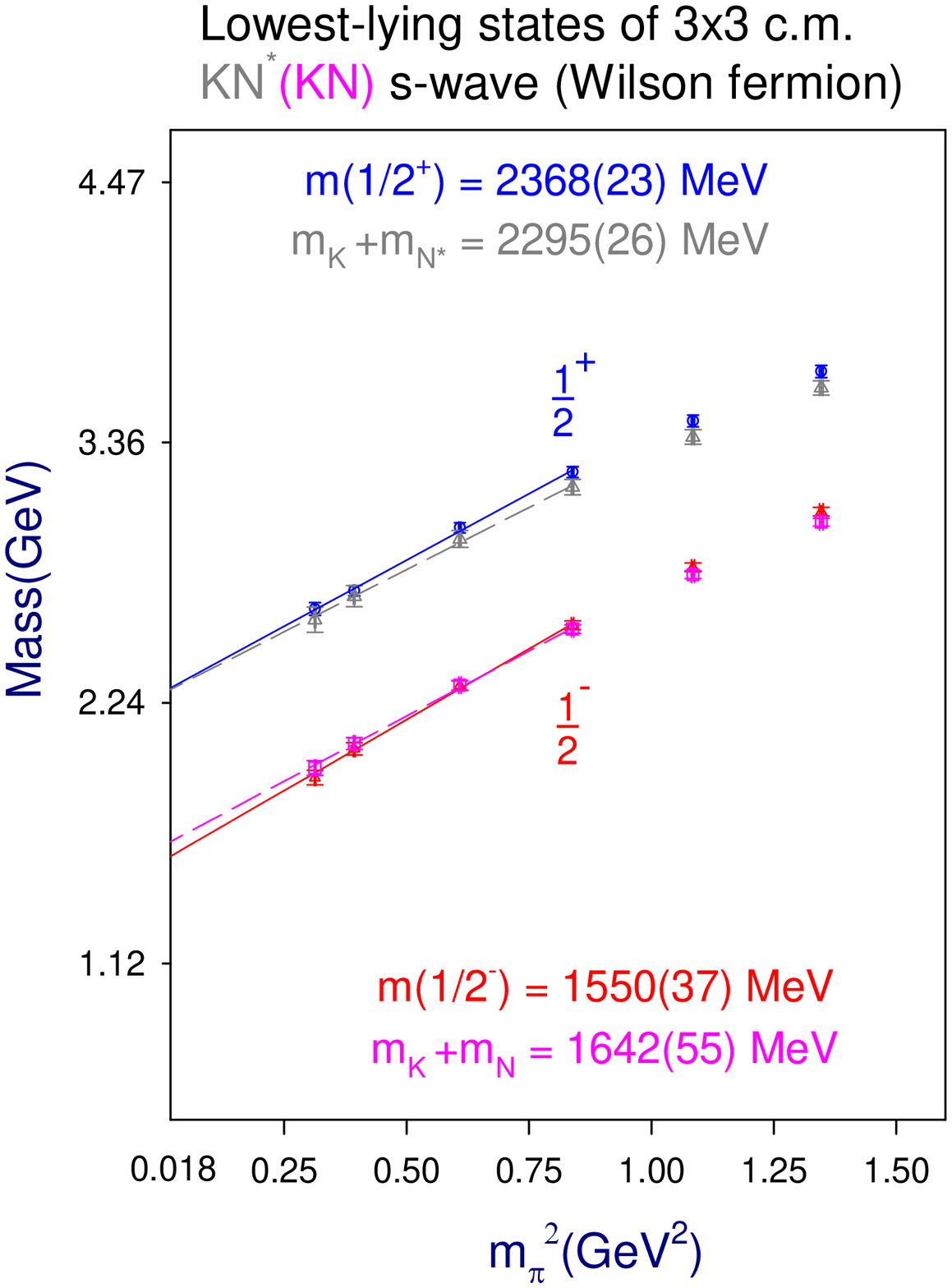}}
\end{tabular}
\vskip -0.8truecm
\caption{The lowest-lying states of the $3\times 3$ correlation
        matrix versus $KN$ and $KN^*$ (s-wave) scattering states,
        for (a) overlap, and (b) Wilson fermions respectively.}
\label{fig:3x3 vs KN KN^*}
\end{center}
\end{figure}

Next we compare the lowest-lying states extracted from the 
$3\times 3$ correlation matrix with the 
$KN$ and $KN^*$ (s-wave) scattering states,
as shown in \Fig{fig:3x3 vs KN KN^*}. 
Naively, assuming the interaction between the Kaon and the Nucleon in 
these (s-wave) scattering states is very weak, then their masses can be 
estimated as $ m_K + m_N $ and $ m_K + m_{N^*} $ respectively.  
For the $ 1/2^-$ state, its mass agrees with 
the $KN$ s-wave, for overlap and Wilson fermions resepctively. 
On the other hand, for the $ 1/2^+$ state, 
it agrees with the $KN^*$ s-wave for Wilson fermion, 
but it is different from the $ KN^* $ s-wave for overlap fermion.  
This reveals an important fact that one cannot identify the 
nature of any excited state by simply comparing its mass with those  
of scattering states.  

\begin{figure}[th]
\begin{center}
\begin{tabular}{@{}c@{}c@{ }c@{}c@{}}
\parbox[b][7.5cm][t]{6.5mm}{(a)} &
\parbox[b][7.5cm][t]{6.5cm}{
        \includegraphics*[height=7cm,width=6cm]{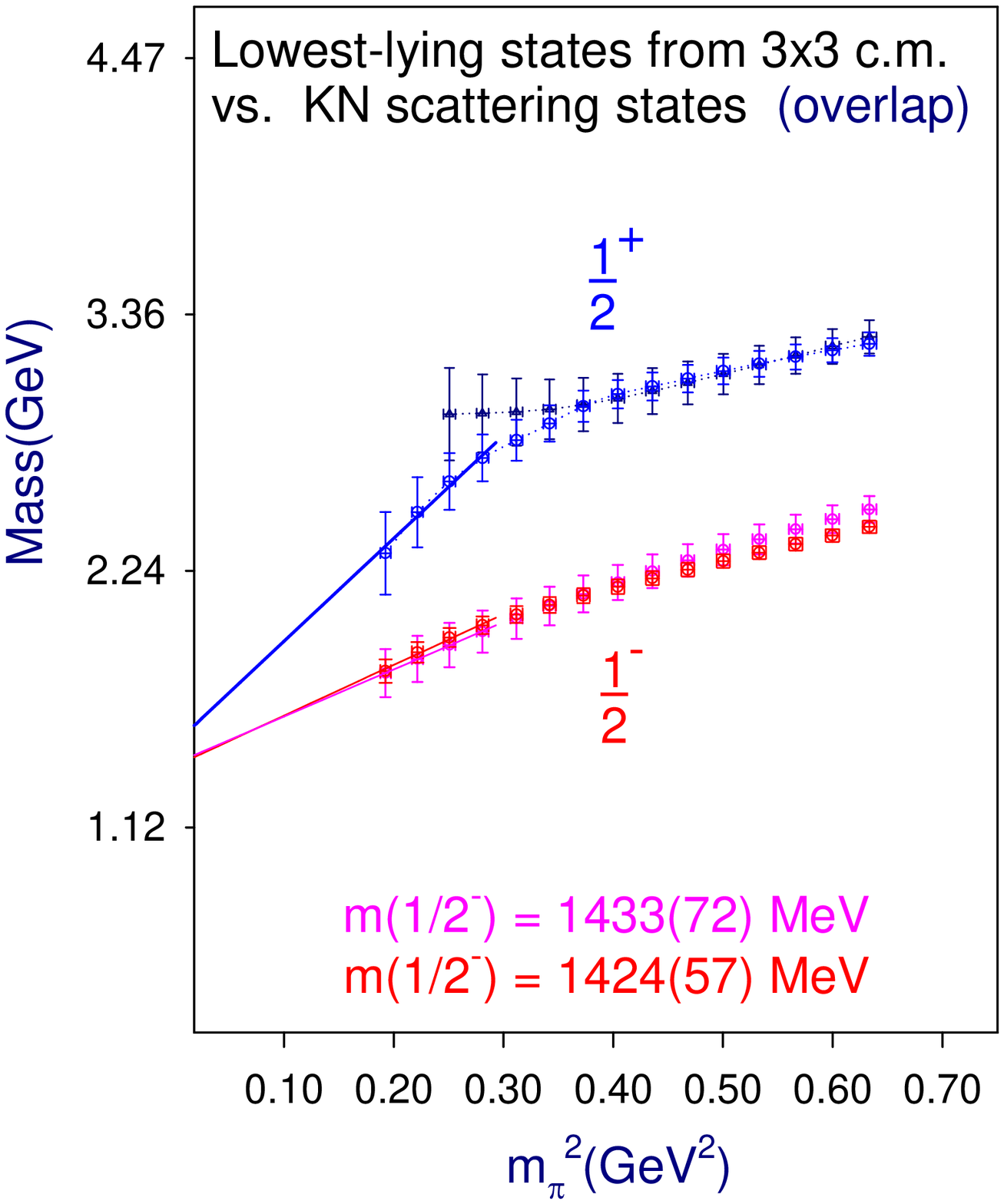}} &
\parbox[b][7.5cm][t]{6.5mm}{(b)} &
\parbox[b][7.5cm][t]{6.5cm}{
        \includegraphics*[height=7cm,width=6cm]{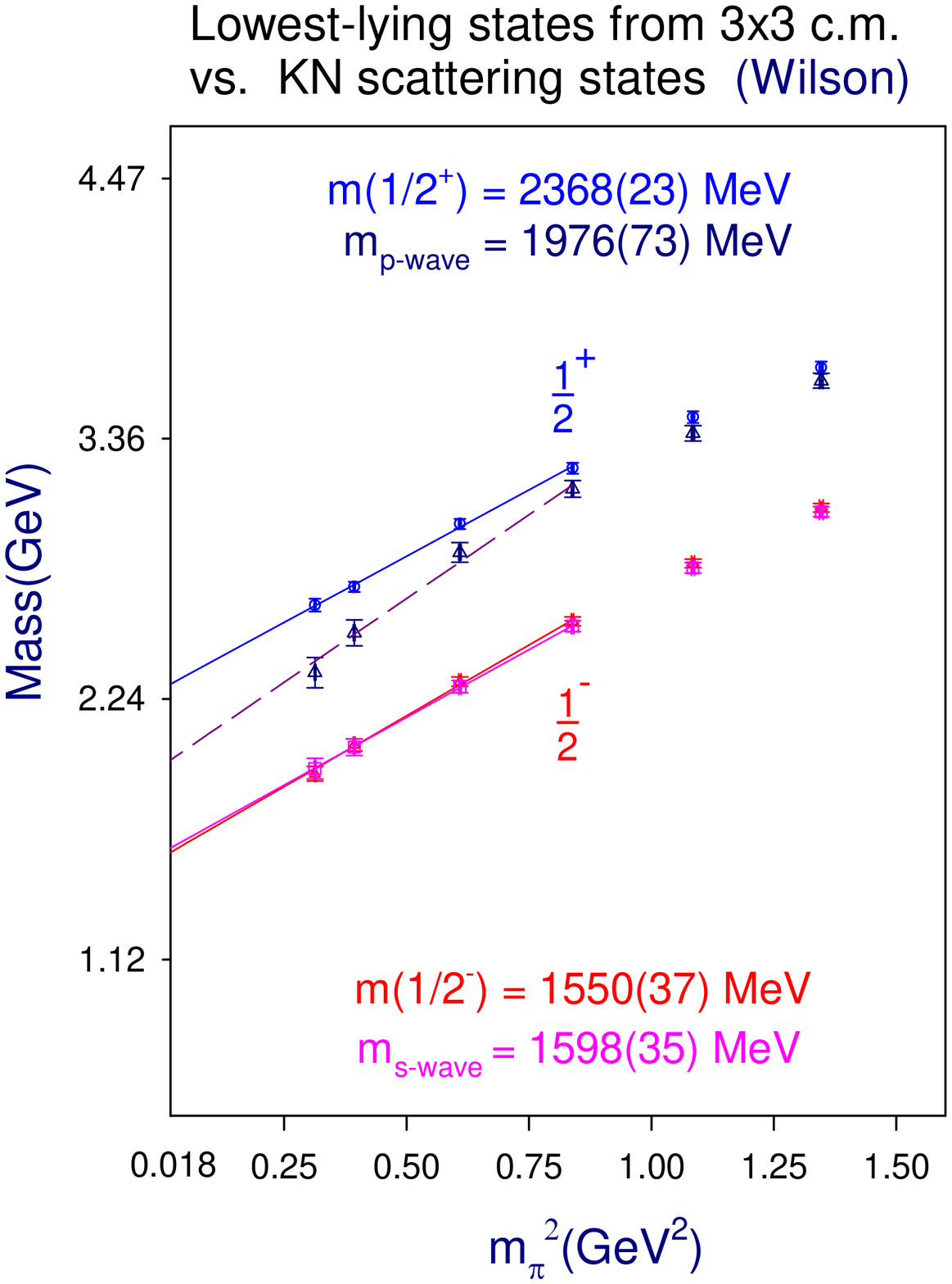}}
\end{tabular}
\vskip -0.8truecm
\caption{The lowest-lying states from $3\times 3$ correlation matrix
         versus $KN$ scattering states,
         for (a) overlap, and (b) Wilson fermions respectively.}
\label{fig:3x3 vs KN-disconnect}
\end{center}
\end{figure}

To obtain a more reliable estimate of the mass spectra of 
$ KN $ scattering states, we measure the correlation function 
of $ KN $ operator without any exchange 
of quarks between the propagators of $ K $ and $ N $, i.e., 
\beq
\label{eq:KN-disconnect}
C^{\pm}_{KN}(t) =
        \Bigl<\sum_{\vec x}\tr\Bigl[\frac{1\mp\gamma_4}{2}
        \left< N(\vec x,t)\bar N(\vec 0,0)\right>_f
        \left< K(\vec x,t)\bar K(\vec 0,0)\right>_f
        \Bigr]\Bigr>_U
\eeq
\begin{figure}[th]
\begin{center}
\begin{tabular}{@{}c@{}c@{ }c@{}c@{}}
\parbox[b][7.5cm][t]{6.5mm}{(a)} &
\parbox[b][7.5cm][t]{6.5cm}{
        \includegraphics*[height=7cm,width=6cm]{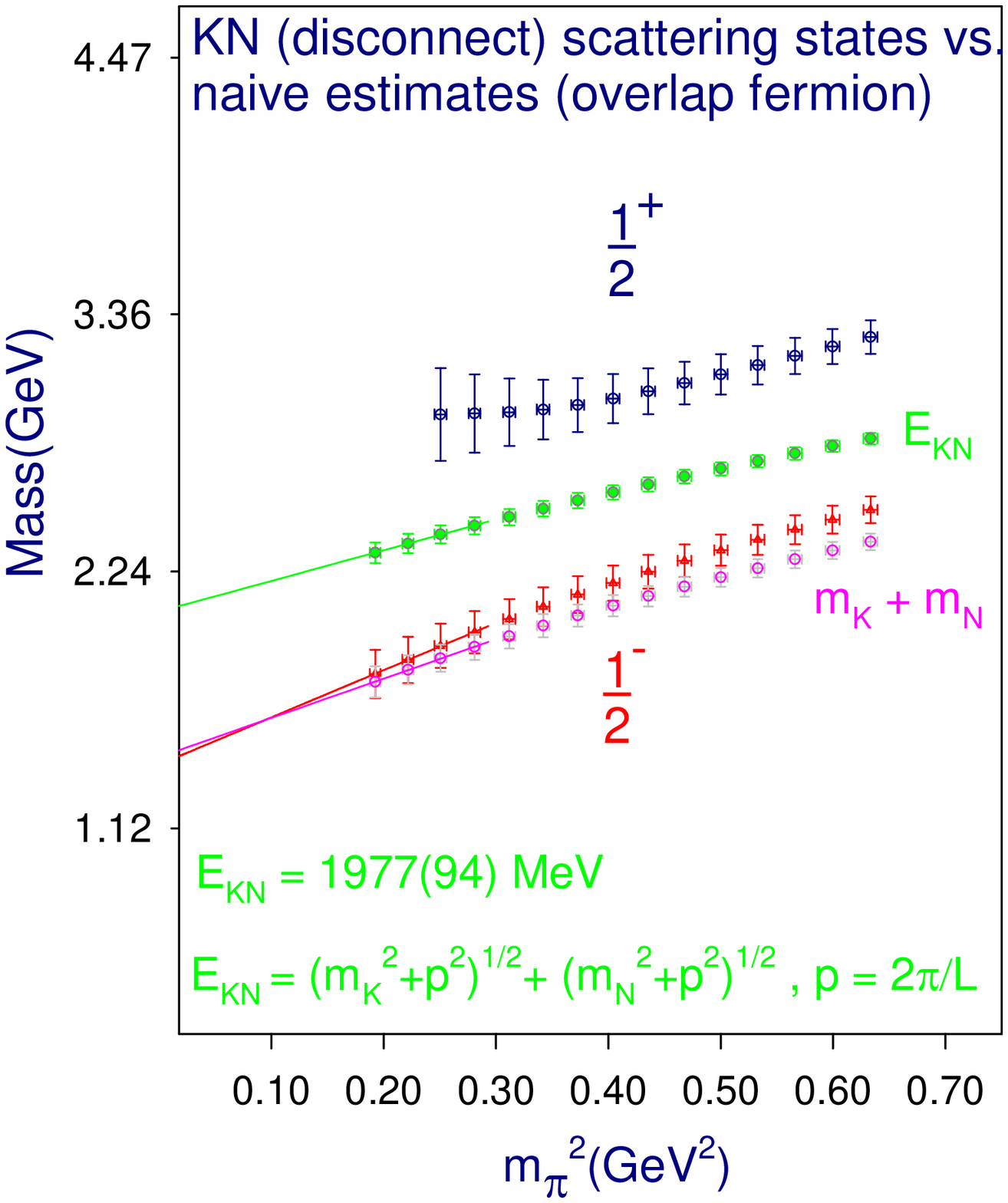}} &
\parbox[b][7.5cm][t]{6.5mm}{(b)} &
\parbox[b][7.5cm][t]{6.5cm}{
        \includegraphics*[height=7cm,width=6cm]{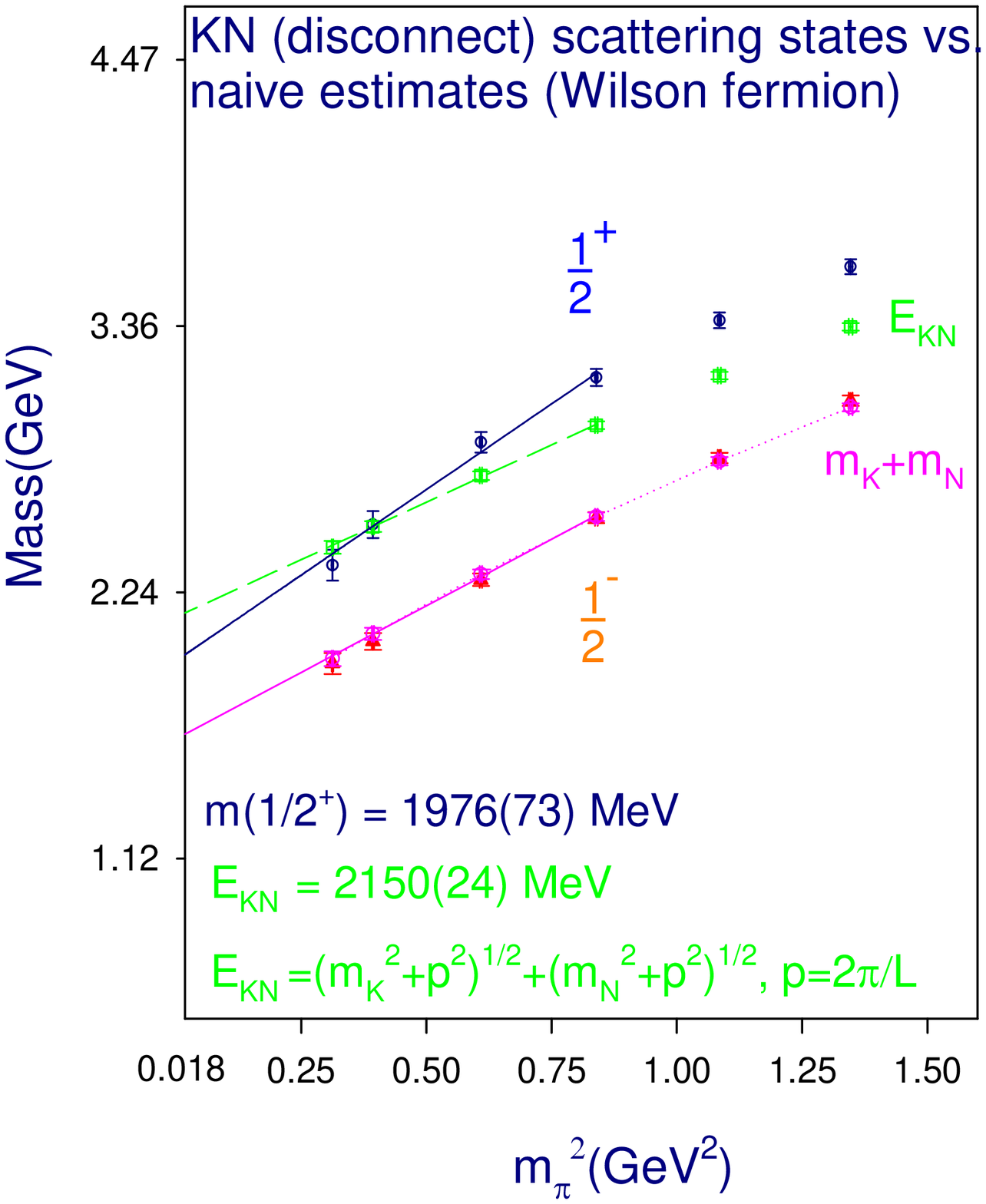}}
\end{tabular}
\vskip -0.8truecm
\caption{The mass spectra of $KN$ (disconnected) scattering states versus
    naive estimates, for (a) overlap, and (b) Wilson fermions respectively.}
\label{fig:KN-disconnect}
\end{center}
\end{figure}
Then we compare its mass spectrum with the naive estimates, 
$ \sqrt{m_K^2 + (2 \pi/L)^2} + \sqrt{m_N^2 + (2 \pi/L)^2} $ (p-wave),
and $ m_K + m_N $ (s-wave), 
as shown in \Fig{fig:KN-disconnect}.  
For the $1/2^-$ state, it agrees with $m_K + m_N$ very well, 
for overlap and Wilson fermions. 
On the other hand, for the $ 1/2^+ $ state, 
it disagrees with the naive estimate for both fermion schemes.  
This suggests that the $KN$ p-wave (in the quenched
approximation) in a finite torus is more complicated than just two free
particles with opposite momenta $\vec p_K = -\vec p_N = 2\pi\hat e_i/L$.
Comparing the mass spectra of the $ KN $ scattering 
states with those from $ 3 \times 3 $ correlation matrix, 
as in \Fig{fig:3x3 vs KN-disconnect}, 
we identify the lowest $ 1/2^-$ state of $\Theta(udud\bar s)$ 
with the $KN$ s-wave, and also rule out the possibility 
that the lowest $ 1/2^+ $ state could be the $ KN $ p-wave.

\begin{figure}[th]
\begin{center}
\begin{tabular}{@{}c@{}c@{ }c@{}c@{}}
\parbox[b][7.5cm][t]{6.5mm}{(a)} &
\parbox[b][7.5cm][t]{6.5cm}{
        \includegraphics*[height=7cm,width=6cm]{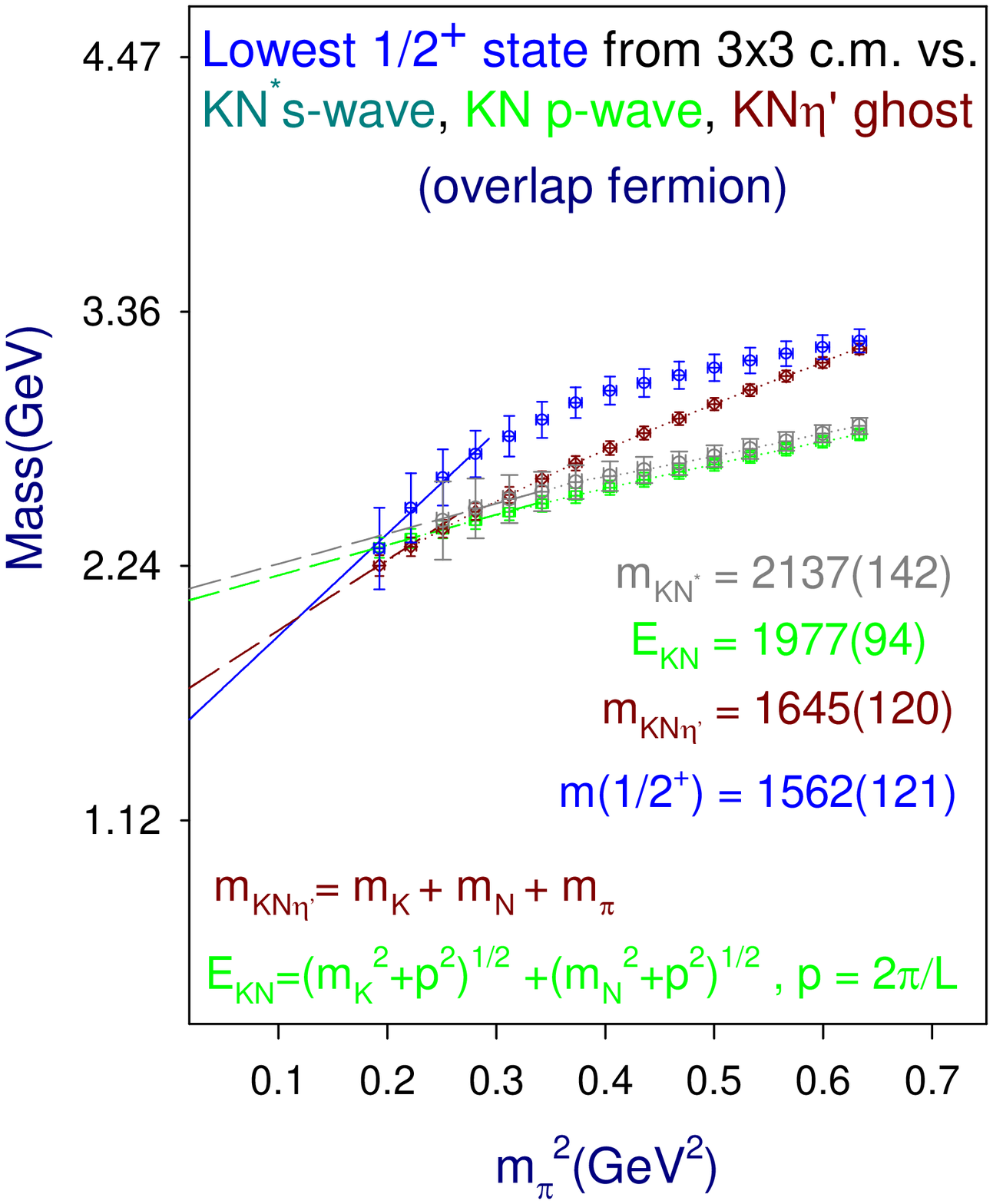}} &
\parbox[b][7.5cm][t]{6.5mm}{(b)} &
\parbox[b][7.5cm][t]{6.5cm}{
        \includegraphics*[height=7cm,width=6cm]{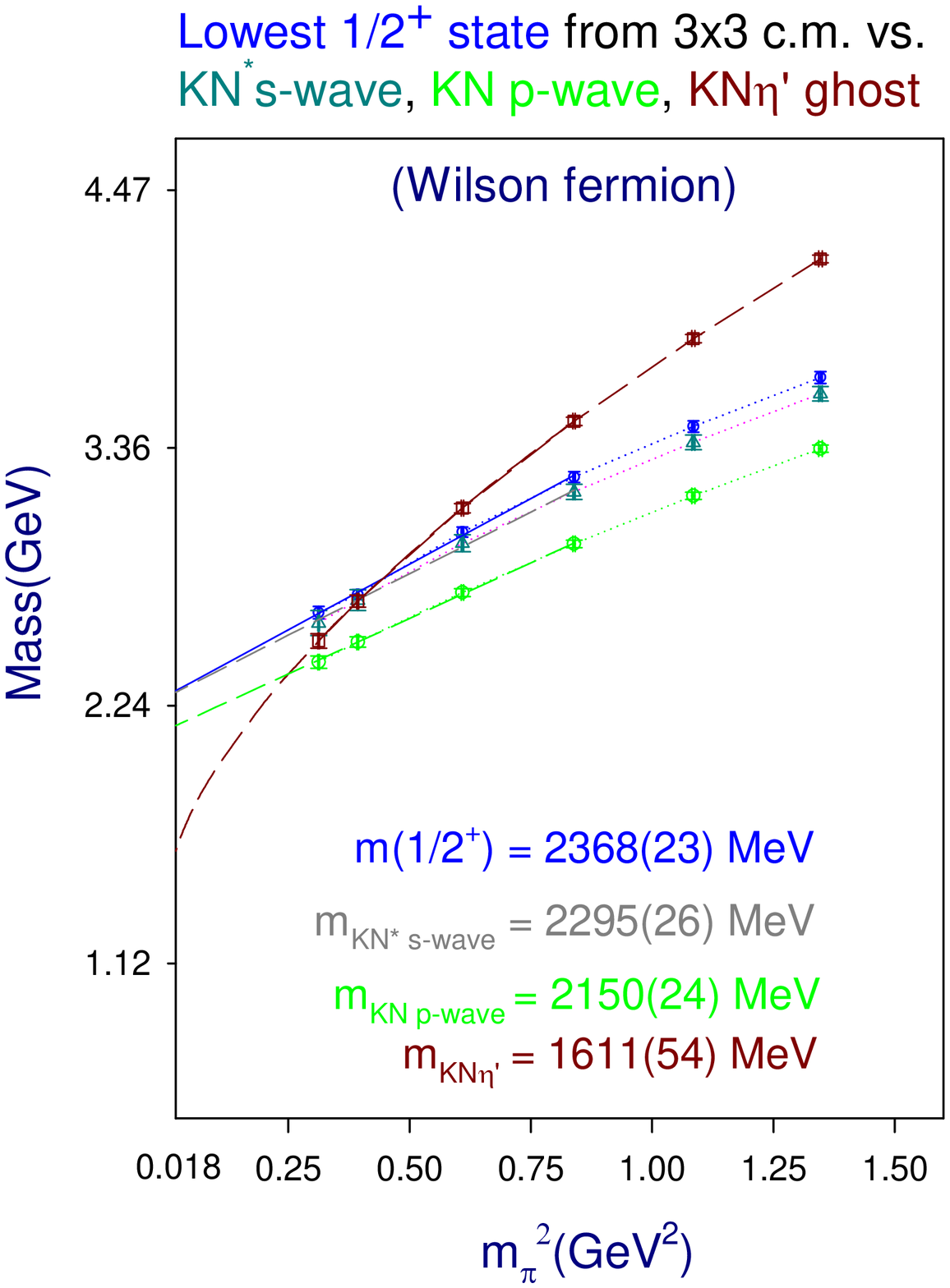}}
\end{tabular}
\vskip -0.8truecm
\caption{The lowest $1/2^+$ state from $3\times 3$ correlation matrix 
         versus $KN^*$ s-wave, $KN$ p-wave, and $KN\eta'$ s-wave,  
         for (a) overlap, and (b) Wilson fermions respectively.}
\label{fig:3x3 1/2^+ vs scattering states}
\end{center}
\end{figure}

Next we consider another hadron scattering state which has the 
same quantum numbers of the $ 1/2^+$ state of $\Theta(udud\bar s)$, 
namely, the s-wave scattering state of $KN\eta'$, where $ \eta' $ 
is an artifact of quenched approximation. 
As shown in \Fig{fig:3x3 1/2^+ vs scattering states}, 
the lowest $ 1/2^+$ state is different from 
the s-wave of $ KN\eta'$, for both fermion schemes. 

\section{Summary and Discussion}

For both fermion schemes, the lowest-lying $ 1/2^- $ state agrees with 
the $ KN $ s-wave scattering state. 
However, for the lowest-lying $ 1/2^+ $ state, 
it is different from any hadron scattering states with the same
quantum numbers (i.e., $KN$ p-wave, $KN\eta'$ s-wave, and $KN^*$ s-wave) 
for the overlap fermion,   
while it seems to agree with $ KN^* $ s-wave for the Wilson fermion.
Obviously, one cannot clarify the nature 
(resonance or scattering state) of any hadronic state 
by simply comparing its mass with those of scattering states.
Further tests (e.g., volume test) are required to resolve the issues 
pertaining to the lowest $ 1/2^+ $ state. In any case, it is vital to 
preserve exact chiral symmetry on the lattice, otherwise, 
the diquark correlations (if any) could not be captured. 
In closing, we note that, so far, none of the exploratory studies 
in lattice QCD \cite{Csikor:2003ng}-\cite{Alexandrou:2005gc}
has reached an unambiguous answer to the question  
{\em whether the spectrum of QCD possesses any resonance 
around $ 1540 $ MeV with $ S = +1 $.}

\end{document}